\documentclass[twocolumn,showpacs,preprintnumbers,amsmath,amssymb,prl,superscriptaddress]{revtex4-1}
\usepackage{graphicx}
\usepackage[caption=false]{subfig}
\usepackage{epstopdf}
\usepackage{dcolumn}
\usepackage{bm}
\usepackage{amsmath}
\usepackage{amssymb}
\usepackage{mathrsfs}
\usepackage{amsfonts}
\usepackage{color}
\usepackage{lipsum}

\begin{document}
\title{Three Flavoured neutrino oscillations and the Leggett Garg Inequality} 

\author{Debashis Gangopadhyay}
\affiliation{Department of Physics, Ramakrishna Mission Vivekananda University, Belur Matth, Howrah, West Bengal, India}

\author{Animesh Sinha Roy}
\affiliation{Department of Physics, Ramakrishna Mission Vivekananda University, Belur Matth, Howrah, West Bengal, India}

\date{\today}


\begin{abstract}
Three flavoured neutrino oscillations are investigated in  the light of the Leggett-Garg inequality. The outline of an experimental proposal is suggested whereby the findings of this investigation may be verified. The results obtained are: 
(a) The maximum violation of the Leggett Garg Inequality (LGI) is $2.17036$ for neutrino path length $L_{1}=140.15 $ Km and $\Delta L=1255.7 $ Km.(b) Presence of the mixing angle  $\theta _{13}$ enhances the maximum violation of LGI by $4.6\%$.(c) The currently known mass hierarchy parameter $\alpha = 0.0305$ increases the the maximum violation of LGI by $3.7\%$. (d)Presence of CP violating phase  parameter enhances the maximum violation of LGI by $0.24\%$, thus providing an \textit{alternative indicator of CP violation} in 3-flavoured neutrino oscillations. 
\end{abstract}

\pacs{42.50.Pq, 42.50 Wk, 07.10.Cm, 42.79 Gn}

\maketitle

\section{Introduction}
The Leggett Garg Inequality (LGI) \cite{LGI} is useful to test the quantumness of a system through successive measurement outcomes at different times on the same system. In our previous work \cite{Gangopadhyay} we showed that two-states neutral kaon oscillations and two-states neutrino oscillations are quantum phenomena by demonstrating that the LGI is violated in both cases.   

Note that the kaon and neutrino cases comprised two different kinds of two state systems. Oscillations between $K^{0}$-$\bar{K}^{0}$ states indicate a decaying two state oscillating quantum system. On the other hand, neutrino oscillations between the two flavour eigenstates $\nu _{e}$ and $\nu _{\mu}$ signify  a conservative two state quantum system. In \cite{Gangopadhyay} for a decaying kaon system, the maximum  violation of LGI  in the presence of CP violation is when the correlator $C = 2.36463$ (defined below in Section 2)  while in the absence of CP violation the LGI violation is maximum when $C=2.36448$. This is significantly smaller than the Tsirelson bound for the LGI in two-states system given by $C_{Tsirelson}=2\sqrt{2}=2.82843$. In case of conservative two-flavour neutrino oscillations the maximum violation of LGI  is  which is when $C =2.76000$. Similar work has also been done in two-states neutrino oscillations \cite{Formaggio}. There the authors have demonstrated how oscillation phenomena can be used to test for violations of the classical bound by performing measurements on an ensemble of neutrinos at distinct energies.

Existence of neutrino mass has been a subject of keen interest over the last fifty years \cite{ponte,gribov}. In 2001 the third generation of neutrinos (tau neutrino) were discovered by the DONUT collaboration \cite{donut}. Exhaustive details regarding various aspects of neutrino masses and oscillations can be found in \cite{gonz,king} and references therein. Therefore, we are motivated to investigate the LGI in the scenario of 3-flavoured neutrino, both without and with CP violation. The effect of CP violation for three flavoured neutrino oscillations may stimulate further investigations in this area. We also consider matter interactions with the neutrino. Here we have analysed the LGI in the context of two small parameters, \textit{viz.} the sine of the  mixing angle $\theta_{13}$, $\sin\theta_{13}<<1$ and the mass hierarchy parameter $\alpha << 1$. Note that the mixing angles are Eulerian angles relating the ($\nu _{e}$, $\nu _{\mu}$, $\nu _{\tau}$) to the mass eigenstates ($\nu _{1}$, $\nu _{2}$, $\nu _{3}$) in the relevant space as shown in fig.1.

\begin{figure}
\resizebox{8.0cm}{5cm}{\includegraphics{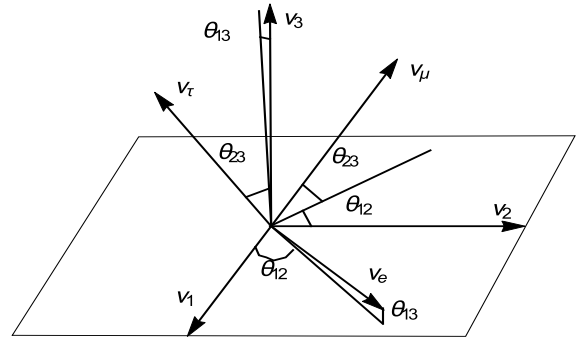}}
\caption{ Neutrino mixing angles without CP violation as Euler angles relating
($\nu _{e}$, $\nu _{\mu}$, $\nu _{\tau}$) to the mass eigenstates ($\nu _{1}$, $\nu _{2}$, $\nu _{3}$).}
\end{figure}

In Section 2, we give a brief introduction of LGI. In Section 3  we discuss the three flavoured neutrino oscillations. In Section 4 the LGI is evaluated and analysed. In Section 5 an  outline is given of  how one can actually experimentally verify the  LGI in three flavoured neutrino oscillations. Section 6 summarises our results. Appendix is in Section 7.

\section{Leggett-Garg Inequality}
Bell's inequality (BI) \cite{Bell} is based on the assumption of \textit{Local Realism} - a intrinsic property of classical physics. Violation of Local Realism signifies  quantum phenomena. BI is a testable algebric inequality constructed from certain combination of correlation functions for the outcomes of an observable quantity measurement on two spatially seperated system at the same instant of time. BI is violated by quantum physics in the present of quantum entanglement between two spatially seperated system, and implies that the quantum world is non-local \cite{aspect}. Later Leggett and Garg \cite{LGI} constructed another algebraic inequality  based on the assumption of \textit{Macrorealism} in terms of the time separated correlation functions corresponding to the successive measurement outcomes at different times on a single system. 

The assumptions underlying the  Leggett-Garg Inequality \cite{LGI} are \textit{Macroscopic Realism} (MR) and  \textit{Noninvasive measurability}(NIM). MR means a macroscopic system during its time evolution, is (at any instant time) in a definite one of the available states. NIM means it is possible in principle to determine which of the states the system is in, without affecting the states itself or the system's subsequent dynamics. These two together constitute \textit{Macrorealism}.

Consider a two state system and an observable quantity $Q(t)$ such that whenever measured it takes values $+1$ or $-1$ for the system is in state 1 or 2 respectively. Further consider a collection of runs starting from identical initial conditions such that in the first set of runs $Q$ is measured at times $t_{1}$ and $t_{2}$; in the second at $t_{2}$ and $t_{3}$; in the third at $t_{3}$ and $t_{4}$; in the fourth at $t_{1}$ and $t_{4}$ ($t_{1} < t_{2} < t_{3} < t_{4}$). From such measurement it is straight forward to determine the temporal correlation function $C_{ij}\equiv \langle Q(t_{i})Q(t_{j})\rangle$ and any physical system obeying asumption of macrorealistic theory gives the Leggett-Garg inequality \cite{LGI}:
\begin{equation}
\label{lgi}
C\equiv C_{12} +C_{23} +C_{34} -C_{14} \leq 2
\end{equation}
A wide range of various quantum system violates the upper bound of LGI and that let the LGI to use it to probe the quantum mechanics (QM) in the macroscopic regime \cite{van, ruskov, gossin, laloy, souza, waldherr, athalye, emar, moreira, budroni, halliwell, lambert, sinha}. A detailed review about LGI can be found in \cite{emary}.

The Legget-Garg Inequlity involves the time parameter whereas the probabilities (given below in Section 3 onwards) are expressed in terms of the base line length parameter $L$. But $L=ct$, $c$ is the velocity of light. So $t$ is automatically present. Now the correlations in time are transcribed into correlations in length.

Consider an $n$-states system. As before, measurements of a macroscopic property $Q$ can yield only two value $\pm 1$, i.e. $Q$  is a dichotomic variable. If some states (say $k$ states where $k < n$) take the value $+1$ then all the remaining $n-k$ states will take the value $-1$. This is no problem because  states with same value of $Q$  may be considered as \textit{microscopically} distinct states with  same macroscopic property $Q$. MR and NIM  
then imply that the system has a definite value of $Q$ at all times and this value 
is independent of previous measurements on the system. Therefore, the bound for Eq. (\ref{lgi}) in macrorealistic theories remains the same.
 
We now consider LGI in 3-states systems with specific attention to the three flavoured neutrino oscillations.

\section{Three flavoured neutrino oscillations}
During propagation neutrinos undergo oscillations among the three flavoured eigenstates $\nu _{e}$, $\nu _{\mu}$ and $\nu _{\tau}$. Consider the standard parameterization of the Pontecorvo-Maki-Nakagawa-Sakata (PMNS) matrix $U$ that mixes the 3 neutrino flavour states \cite{hagi, beringer} :
\begin{eqnarray}
&&U=\nonumber\\
&&\small\begin{pmatrix}
c_{12}c_{13} & s_{12}c_{13} & s_{13}e^{-i\delta _{CP}} \\
-s_{12}c_{23} -c_{12}s_{13}s_{23}e^{i\delta _{CP}} & c_{12}c_{23}-s_{12}s_{13}s_{23}e^{i\delta _{CP}} & c_{13}s_{23} \\
s_{12}s_{23} -c_{12}s_{13}c_{23}e^{i\delta _{CP}} & -c_{12}s_{23}-s_{12}s_{13}c_{23}e^{i\delta _{CP}} & c_{13}c_{23}
\end{pmatrix}\nonumber\\
\end{eqnarray}
 where $\theta _{ij}$ are the mixing angles, $c_{ij}\equiv \cos \theta _{ij}$, $s_{ij}\equiv \sin \theta _{ij}$ and $\delta _{CP}$ is the Dirac-type CP-violating phase. If $P_{\alpha\beta} \equiv P(\nu _{\alpha}\rightarrow \nu _{\beta})$ be  the  transition probability  from one neutrino flavour $\alpha $ to another flavour $\beta$, then 
in general the functional dependence of  $P_{\alpha\beta}$ is
\begin{eqnarray}
P_{\alpha\beta}=P_{\alpha\beta}(\Delta m_{21}^{2},\Delta m_{31}^{2},\theta _{12},\theta_{13},\theta _{23},\delta _{CP},E,L,V(x))\nonumber\\
\end{eqnarray}
where $\alpha ,\beta\equiv e,\mu ,\tau$ .
Here $\Delta m_{ij}^{2}\equiv m_{i}^{2}-m_{j}^{2}$ with 
$m_{i}$ being the mass of the $i-$th species. E is the neutrino energy, L is the baseline length, and V (x) is the matter-induced effective potential, $x \in [0, L]$ is the coordinate along the neutrino path. 

$\Delta m_{ij}^{2}$, $\theta_{ij}$'s and $\delta _{CP}$ are experimentally independent fundamental parameters. On the other hand $E$, $L$ and $V$ vary from experiment to experiment.

In \cite{Akhmedov} complete sets of series expansion formulas for neutrino oscillation probabilities in matter of constant density have been calculated taking into account the  three flavours. We will be considering the neutrino energies of the order of $1$ GeV. Therefore we consider the appropriate double expansion given in   \cite{Akhmedov} upto the second order in both mass hierarchy parameter $\alpha \equiv \frac{\Delta m_{21}^{2}}{\Delta m_{31}^{2}}$ and $s_{13}$. 

Let us start with an electron neutrino beam at time $t=0$, i.e. $L=0$. After time $t$, i.e. distance $L=ct$, the probability of finding $\nu _{e}$, $\nu _{\mu}$ and $\nu _{\tau}$ are respectively \cite{Akhmedov}
 \begin{eqnarray}
 \label{pro-e}
 P_{\nu _{e}}=&&1-\alpha ^{2}\sin ^{2}2\theta_{12}\frac{\sin ^{2}\big(\frac{VL}{2}\big)}{\Big(\frac{2EV}{\Delta m^{2}_{31}}\Big)^{2}}\nonumber\\
 &&-4s^{2}_{13}\frac{\sin ^{2}\Big\{\Big(\frac{2EV}{\Delta m^{2}_{31}}-1\Big)\frac{\Delta m^{2}_{31}L}{4E}\Big\}}{\Big(\frac{2EV}{\Delta m^{2}_{31}}-1\Big)^{2}}
 \end{eqnarray}
\begin{eqnarray}
\label{pro-mu}
&&P_{\nu _{\mu}}=\alpha ^{2}\sin ^{2}2\theta_{12}c^{2}_{23}\frac{\sin ^{2}\big(\frac{VL}{2}\big)}{\Big(\frac{2EV}{\Delta m^{2}_{31}}\Big)^{2}}+4s^{2}_{13}s^{2}_{23}\times \nonumber\\
&&\frac{\sin ^{2}\Big\{\Big(\frac{2EV}{\Delta m^{2}_{31}}-1\Big)\frac{\Delta m^{2}_{31}L}{4E}\Big\}}{\Big(\frac{2EV}{\Delta m^{2}_{31}}-1\Big)^{2}}+2\alpha s_{13}\sin 2\theta_{12}\sin 2\theta_{23}\nonumber\\
&&\cos (\frac{\Delta m^{2}_{31}L}{4E} -\delta _{CP})\frac{\sin \big(\frac{VL}{2}\big)}{\Big(\frac{2EV}{\Delta m^{2}_{31}}\Big)}\frac{\sin \Big\{\Big(\frac{2EV}{\Delta m^{2}_{31}}-1\Big)\frac{\Delta m^{2}_{31}L}{4E}\Big\}}{\Big(\frac{2EV}{\Delta m^{2}_{31}}-1\Big)}\nonumber\\
\end{eqnarray}
\begin{eqnarray}
\label{pro-tau}
&&P_{\nu _{\tau}}=\alpha ^{2}\sin ^{2}2\theta_{12}s^{2}_{23}\frac{\sin ^{2}\big(\frac{VL}{2}\big)}{\Big(\frac{2EV}{\Delta m^{2}_{31}}\Big)^{2}}+4s^{2}_{13}c^{2}_{23}\times \nonumber\\
&&\frac{\sin ^{2}\Big\{\Big(\frac{2EV}{\Delta m^{2}_{31}}-1\Big)\frac{\Delta m^{2}_{31}L}{4E}\Big\}}{\Big(\frac{2EV}{\Delta m^{2}_{31}}-1\Big)^{2}}-2\alpha s_{13}\sin 2\theta_{12}\sin 2\theta_{23}\nonumber\\
&&\cos (\frac{\Delta m^{2}_{31}L}{4E} -\delta _{CP})\frac{\sin \big(\frac{VL}{2}\big)}{\Big(\frac{2EV}{\Delta m^{2}_{31}}\Big)}\frac{\sin \Big\{\Big(\frac{2EV}{\Delta m^{2}_{31}}-1\Big)\frac{\Delta m^{2}_{31}L}{4E}\Big\}}{\Big(\frac{2EV}{\Delta m^{2}_{31}}-1\Big)}\nonumber\\
\end{eqnarray}

 So it is easy to say that after travelling the distance $L$ probability of obtaining $\nu _{e}$, $\nu _{\mu}$ and $\nu _{\tau}$ are given by equations (\ref{pro-e}), (\ref{pro-mu}) and (\ref{pro-tau}) respectively. The join probability of finding the neutrino with flavours $\nu _{e}$ and $\nu _{\mu}$ after travelling respective distances $L_{1}$ and $L_{2}$ $(L_{2}>L_{1})$ is then
\begin{eqnarray}
\label{joint-pro}
&&P_{\nu _{e},\nu_{\mu}}(L_{1},L_{2})=\Bigg[1-\alpha ^{2}\sin ^{2}2\theta_{12}\frac{\sin ^{2}\big(\frac{VL_{1}}{2}\big)}{\Big(\frac{2EV}{\Delta m^{2}_{31}}\Big)^{2}}-4s^{2}_{13}\times\nonumber\\
&&\frac{\sin ^{2}\Big\{\Big(\frac{2EV}{\Delta m^{2}_{31}}-1\Big)\frac{\Delta m^{2}_{31}L_{1}}{4E}\Big\}}{\Big(\frac{2EV}{\Delta m^{2}_{31}}-1\Big)^{2}}\Bigg]\Bigg[\alpha ^{2}\sin ^{2}2\theta_{12}c^{2}_{23}\frac{\sin ^{2}\big(\frac{V(L_{2}-L_{1})}{2}\big)}{\Big(\frac{2EV}{\Delta m^{2}_{31}}\Big)^{2}}\nonumber\\
&&+4s^{2}_{13}s^{2}_{23}\frac{\sin ^{2}\Big\{\Big(\frac{2EV}{\Delta m^{2}_{31}}-1\Big)\frac{\Delta m^{2}_{31}(L_{2}-L_{1})}{4E}\Big\}}{\Big(\frac{2EV}{\Delta m^{2}_{31}}-1\Big)^{2}}+2\alpha s_{13}\sin 2\theta_{12}\nonumber\\
&&\sin 2\theta_{23}\cos \Big(\frac{\Delta m^{2}_{31}(L_{2}-L_{1})}{4E} -\delta _{CP}\Big)\frac{\sin \big(\frac{V(L_{2}-L_{1})}{2}\big)}{\Big(\frac{2EV}{\Delta m^{2}_{31}}\Big)}\nonumber\\
&&\frac{\sin \Big\{\Big(\frac{2EV}{\Delta m^{2}_{31}}-1\Big)\frac{\Delta m^{2}_{31}(L_{2}-L_{1})}{4E}\Big\}}{\Big(\frac{2EV}{\Delta m^{2}_{31}}-1\Big)}\Bigg].
\end{eqnarray}
In similar way one can calculate the other $8$ joint probabilities.

\section{Evaluating and analysing LGI for 3 flavours of neutrino}
In the three flavoured neutrino oscillations, we assume that the dichotomic observable $Q$ takes the value $+1$ when the system to be found in the electron neutrino flavour state $\nu _{e}$. $Q$ takes the value $-1$ if the system is found in any one of the muon neutrino $\nu _{\mu}$ or tau neutrino $\nu _{\tau}$ states. Then the correlation function $C_{12}$ can be evaluated by using all the $9$ joint probabilities as 
\begin{eqnarray}
C_{12} &&=\langle Q(L_{1})Q(L_{2})\rangle\nonumber\\
&&= P_{\nu _{e},\nu_{e}}(L_{1},L_{2})-P_{\nu _{e},\nu_{\mu}}(L_{1},L_{2})-P_{\nu _{e},\nu_{\tau}}(L_{1},L_{2})\nonumber\\
&&-P_{\nu _{\mu},\nu_{e}}(L_{1},L_{2})+P_{\nu _{\mu},\nu_{\mu}}(L_{1},L_{2})+P_{\nu _{\mu},\nu_{\tau}}(L_{1},L_{2})\nonumber\\
&&-P_{\nu _{\tau},\nu_{e}}(L_{1},L_{2})+P_{\nu _{\tau},\nu_{\mu}}(L_{1},L_{2})+P_{\nu _{\tau},\nu_{\tau}}(L_{1},L_{2})\nonumber\\
\end{eqnarray}
The details expression of the correlation function $C_{12}$ is given in the Appendix. An interesting point in the expression of $C_{12}$ is that for neutrino beam with given energy the correlation $C_{12}$ show dependence on $L_{1}$ as well as the spacial separation $(L_{2}-L_{1})$. It is also important to note that in the case of two flavoured neutrino oscillations the correlation function depend only on the spacial seperation $(L_{2}-L_{1})$ \cite{Gangopadhyay}. The other correlation functions, $C_{23}$, $C_{34}$ and $C_{14}$ can be calculated in the same way and they also give similar feature. Next one can evaluate the correlation function $C$ defined in the Eq.(\ref{lgi}) in order to study the maximum violation of LGI for three flavoured neutrino oscillations. Varying different choices of spacial separations it is found that the maximum value of the correlation function $C$ is attained essentially when all the spatial separations are taken to be same, \textit{i.e.} $(L_{4}-L_{3})=(L_{3}-L_{2})=(L_{2}-L_{1})=\Delta L$ and the correlation function $C$ under this condition depends on $\Delta L$ and $L_{1}$. 

We have calculated the maximum value of $C$ using latest experimentally determined values given in \cite{Garcia}. These are 
 $\Delta m^{2}_{21}=7.50\times 10^{-5} eV^{2}$, $\Delta m^{2}_{31}=2.457\times 10^{-3} eV^{2}$, $\theta _{12}=33.48^{\circ}$, $\theta _{23}=42.3^{\circ}$, $\theta _{13}=8.50^{\circ}$, $\delta _{CP}=306^{\circ}$. Here the potential \cite{Akhmedov} $V=7.56\times 10^{-14} \Big(\frac{\rho}{g/cm^{3}}\Big)Y_{e} eV$, where $\rho$ is matter density along the neutrino path and $Y_{e}$ is the number of electron per nucleon. For terrestrial matter  $Y_{e}\simeq 0.5$ \cite{Akhmedov}. For practical purposes it is a very good approximation to assume $\rho$ to be constant \cite{Nicolaidis, Liu, Freund}. Typical value of matter density is $\rho =3 g/cm^{3}$ \cite{Akhmedov}. So the potential $V$ takes the value $V=11.34\times 10^{-14} eV$. Here we consider the energy of neutrino to be $1 GeV$. Further we consider various choices of $L_{1}$ and $\Delta L$ and found that the maximum value of $C$ reach to 2.17036 for $L_{1}=140.15 $ Km and $\Delta L=1255.7 $ Km. It is very important to note that the maximum QM violation of LGI in this case is significantly smaller than the maximum QM value of $C$ we calculted \cite{Gangopadhyay} in the case of two flavoured neutrino oscillation which was $2.76$. For the given value of $L_{1}=140.15$ Km, the variation of the quantity $C$ with $\Delta L$ is shown in the fig.2.

\begin{figure}
\resizebox{8.0cm}{5cm}{\includegraphics{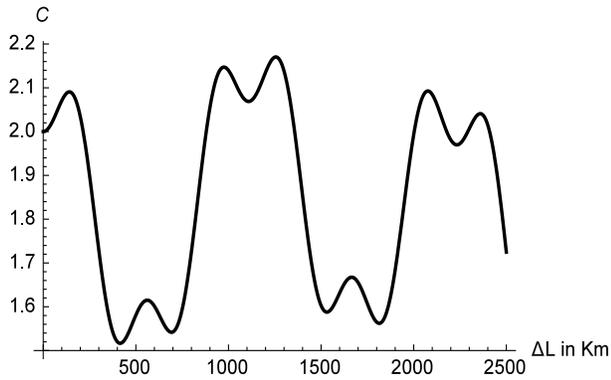}}
\caption{ Correlation $C$ as a function of $\Delta L$ in Km for $L_{1}=140.15$ Km. $C$ attains its maximum value 2.17036 at $\Delta L=1255.7 $ Km. }
\end{figure}

  Now we will investigate how the mixing angle $\theta _{13}$ affect the maximum value of the quantity $C$. If we put $\theta _{13}=0$, the maximum value of $C$ becomes $C=2.07762$ for $L_{1}=638$ Km and $\Delta L=1376.34$ Km. This  is much lower than the actual value $(2.17036)$ when $\theta _{13}\neq 0$.  This value still belongs to the quantum domain because it is lager than $2$. So the presence of the mixing angle $\theta _{13}$ in three flavour neutrino oscillations enhance the maximum violation of LGI by the amount $0.09274$. If we increase $\theta _{13}$ from zero degree we see that the maximum value of the quantity $C$ also  increases. This means increasing the mixing angle $\theta _{13}$ also increases the quantumness of the three flavoured neutrino oscillation. The variation of $C$ with $\theta _{13}$ is shown in the fig.3 below
\begin{figure}
\resizebox{9.0cm}{13cm}{\includegraphics{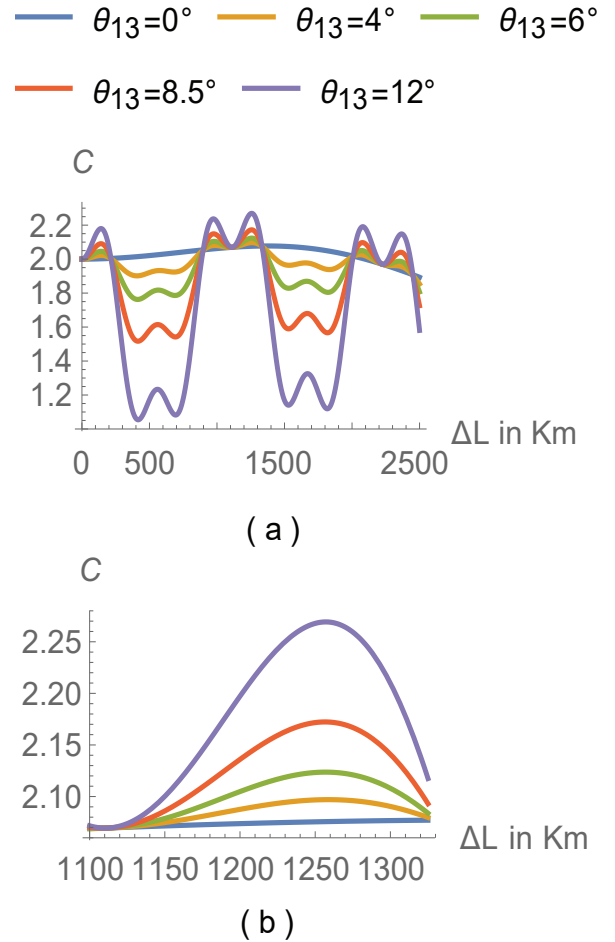}}
\caption{ (a) $C$ versus $\Delta L$ in Km for different values of the mixing angle parameter $\theta _{13}$. Here $L_{1}=140.15$ Km. Blue$\sim 0^{\circ}$, orange $\sim 4^{\circ}$, green$\sim 6^{\circ}$, red$\sim 8.5^{\circ}$(actual experimentally measured value),violate $\sim 12^{\circ}$. (b) We focus around the region where the value of $C$ is maximum. the maximum value of $C$ increase with the increase of the value of the mixing angle $\theta _{13}$}
\end{figure}

One of the key parameter in three flavour neutrino oscillation is the \textit{small} mass hierarchy parameter $\alpha \equiv \frac{\Delta m_{21}^{2}}{\Delta m_{31}^{2}}$. In this section we investigate the dependence of the quantity $C$ with the small mass hierarchy parameter $\alpha$. If we put $\alpha = 0 $, \textit{i.e.} $m_{1}= m_{2}$ the maximum value of the quantity $C$ becomes $2.09606$ for $\Delta L=1252.74$. It is interesting to note that although now $m_{1}= m_{2}$ the maximum bound of the quantity $C$ is greater than $2$, i.e., we are still in the quantum domain. For two state neutrino oscillations, \cite{Gangopadhyay} the condition $m_{1}= m_{2}$ implies that the maximum value of $C$ is $2$, i.e. \textit{one is in the classical domain.!}. This is logical because this means there is only one neutrino mass and so there cannot be any oscillations. However, for three state neutrino oscillations there are three neutrino masses and if two of them become equal then also there will exist possibility of neutrino oscillations because now there are effectively two masses. In the present case of three state neutrino oscillation the presence of the non zero value of $\alpha$ increase the the maximum value of the quantity $C$ which is shown in the fig.4. In the fig.4 Blue, orange, green and red color graphs are the behaviour of the quantity $C$ for the value of $\alpha =$ $0$, $0.01$, $0.0305$(actual experimentally measured value) and $0.06$. For the present experimentally measured value of $\alpha$ the maximum value of the quantity $C$ increase about $3.7\%$. So the presence of non zero $\alpha$ increase the quantumness in the case of three flavor neutrino oscillation.

\begin{figure}
\resizebox{8.7cm}{15cm}{\includegraphics{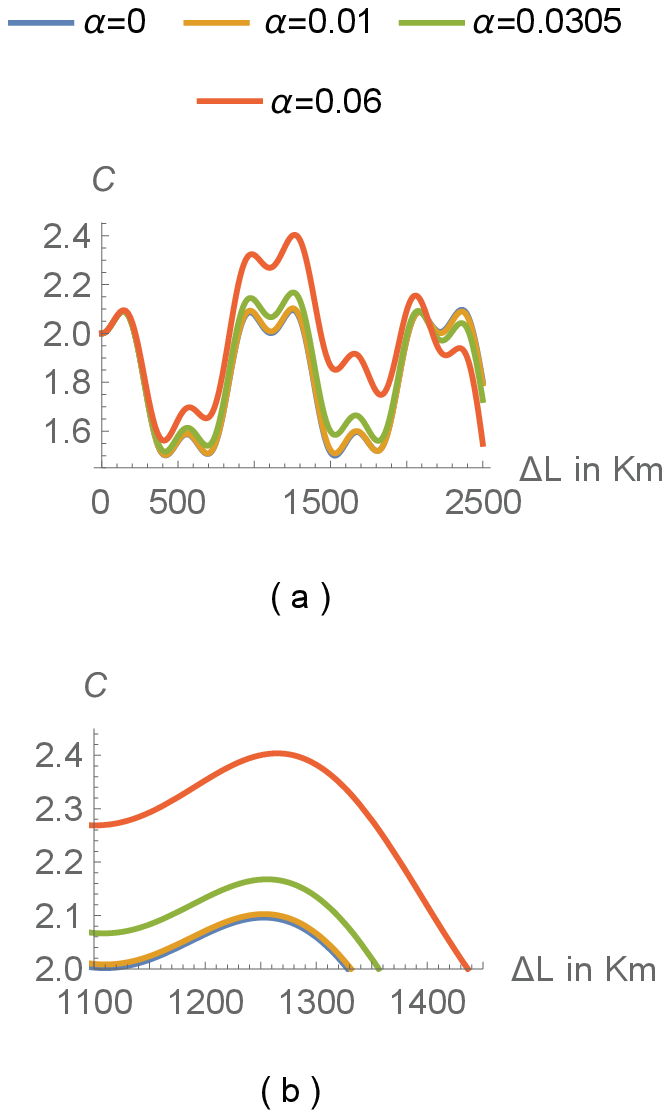}}
\caption{Behaviour of the quantity $C$ with the variation of $\Delta L$ in Km for different values of the mass hierarchy parameter $\alpha$ is shown in the fig.4(a). Here $L_{1}=140.15$ Km. Blue, orange, green and red color graphs are the behaviour of the quantity $C$ for the value of $\alpha =$ $0$, $0.01$, $0.0305$(actual experimentally measured value) and $0.06$. In the fig.4(b) we focus around the region where the value of the quantity $C$ is maximum. From Fig.4 we observe that the maximum value of $C$ increase with the increase of the value of the mass hierarchy parameter $\alpha$}
\end{figure}

Next we investigate the effect of the CP violating phase $\delta _{CP}$ parameter in the maximum value of the quantity $C$. If we ignore the CP violating phase $\delta _{CP}$ parameter in the expression of the quantity $C$ we observe that the maximum value of $C$ reduces to $2.16553$ for $L_{1}=140.15$ Km and $\Delta L=1253.8$ Km. So presence of CP violating phase $\delta _{CP}$ parameter actually enhance the maximum violation of LGI by an amount $0.00483$, a significant enhancement. That means the effect of the CP violation actually enhance the quantum ness of the three flavor neutrion oscillation system. It is worth to mention that in the case of neutral kaon oscillation the presence of the CP violation enhance the maximum violation of LGI by an amount $0.00015$ \cite{Gangopadhyay} which is $0.008\%$ enhancement whereas here the effect of CP violation increment of maximum violation of LGI is $0.24\%$. So when LGI is concern the effect of the CP violation is much more in three flavoured neutrino oscillation compared to neutral kaon system. In the fig.5 we focus around the region where the the quantity $C$ takes its maximum value both with and without CP violation.

\begin{figure}
\resizebox{8.0cm}{5.5cm}{\includegraphics{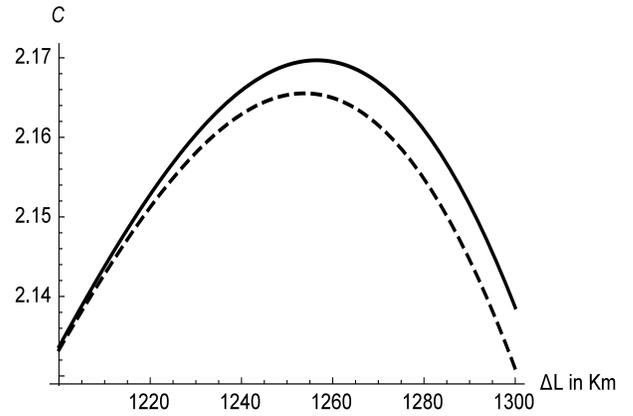}}
\caption{Variation of the quantity $C$ as a function of $\Delta L$ in Km  with and without CP violation  for $L_{1}=140.15$ Km is shown in the Fig.3. Here we focus around the region where the the quantity $C$ takes its maximum value both with and without CP violation. The solid curve is the behaviour of $C$ including CP violation and the dashed curve is the behaviour of $C$ without CP violation. Fig.5 tells that the presence of CP violation enhance the maximum QM violation of LGI.}
\end{figure}

\section{A proposal for Experimental verification }
To test experimentally the maximum violation of LGI for three flavoured neutrino oscillations the first thing necessary is the determination of the correlation function $C_{12}$. For this the  observable quantity $Q$ has to be measured at two different times $t_{1}$ and $t_{2}$ ($t_{2} > t_{1}$) or equivalently at two different base line lengths $L_{1}$ and $L_{2 }$ where 
$L_{2} > L_{1}$. As  already mentioned $Q$ takes the value $+1$ when the system is found in electron-neutrino flavour state. Otherwise $Q$ takes the value $-1$. So 
\begin{eqnarray}
C_{12} = &&
\mathcal{P}_{++}(L_{1}, L_{2})-\mathcal{P}_{+-}(L_{1},L_{ 2})-\mathcal{P}_{-+}(L_{1}, L_{2})\nonumber\\
&&+\mathcal{P}_{--}(L_{1}, L_{2})
\end{eqnarray}
 
where $\mathcal{P}_{++}(L_{1}, L_{2})={P}_{\nu _{e},\nu _{e}}(L_{1}, L_{2})$ is the joint probability of finding the system in the electron neutrino flavor state at both the distances $L_{1}$ and $L_{2}$. Similar arguments hold for the other $3$ joint probabilities:
$$\mathcal{P}_{+-}(L_{1}, L_{2})={P}_{\nu _{e},\nu _{\mu}}(L_{1}, L_{2})+{P}_{\nu _{e},\nu _{\tau}}(L_{1}, L_{2})$$
$$\mathcal{P}_{-+}(L_{1}, L_{2})={P}_{\nu _{\mu},\nu _{e}}(L_{1}, L_{2})+{P}_{\nu _{\tau},\nu _{e}}(L_{1}, L_{2})$$
$$\mathcal{P}_{--}(L_{1}, L_{2})={P}_{\nu _{\mu},\nu _{\mu}}(L_{1}, L_{2})+{P}_{\nu _{\mu},\nu _{\tau}}(L_{1}, L_{2})$$
$$+{P}_{\nu _{\tau},\nu _{\mu}}(L_{1}, L_{2})+{P}_{\nu _{\tau},\nu _{\tau}}(L_{1}, L_{2})$$
\textit{Note that the scripted probabilities $\mathcal{P}$ are the ones that are actually measured.} These are  related to the theoretically calculated unscripted probabilities as shown above. This is necessitated by the fact that here more than one state can have the same value for the dichotomic variable $Q$.

It is to be noted that to experimentally verify the maximum violation of LGI the first measurement of $Q$ at length $L_{1}$ must satisfy NIM. Otherwise measurement proceess will destroy the state of the system and measurement of $Q$ at the later length $L_{2}$ will be meaningless as the state has already been disturbed. This (NIM in the first measurement) can be ensured using the negative result measurement (NRM) \cite{home} as follows. Let the measuring set-up be arranged so that if the probe is triggered, $Q(L_{1})=+1$, while if it is not triggered , $Q(L_{1})= -1$. This ensures that while the untriggered probe provides information about the value of Q, there is no interaction occurring between the probe and the measured particle. So  NIM is satisfied. Now use only the results of  untriggered  runs for which $Q(L_{1})=-1$. Follow this by the measurement of $Q$ at $L_{2}$. These results can be used for determining the joint probabilities $P_{-+}(L_{1}, L_{2})$ and $P_{--}(L_{1}, L_{2})$. Similarly, for determining the other two joint probabilities $P_{+-}(L_{1}, L_{2})$ and $P_{++}(L_{1}, L_{2})$ occurring in $C_{12}$, the measuring set-up can be inverted so that a value of $Q(L_{1})=-1$ triggers the probe, while for $Q(L_{1})=+1$, it does not. In this way, one can determine $C_{12}$ and all the two-time correlation functions occurring in LGI by ensuring NIM through the use of the NRM procedure for the first measurement of any pair. Then one can calculate the total correlation $C$ using eqn.(\ref{lgi}) and experimentally verify our results about the maximum violation of LGI in the case of three flavour neutrino oscillation.

\section{Concluding Remarks}
In this work  we have investigated the violation of the LGI in the case of $3$-flavoured neutrino oscillations. Our findings are as follows : 

(1) The maximum value of the correlation  $C$ is $2.17036$ for $L_{1}=140.15 $ Km and $\Delta L=1255.7 $ Km.

(2)The violation of the classical bound of  $C$ given by LGI for three flavour neutrino oscillation is $8.5\%$.  Note that in the case of $2$ flavour neutrino oscillations \cite{Gangopadhyay} this violation was $38\%$. So the maximum violation of LGI in case of three flavour neutrino oscillations is \textit{significantly lower than the maximum violation for the two state neutrino oscillation}.

(3)If we put $\theta _{13}=0$, the maximum value of $C$ is $2.07762$ for $L_{1}=638$ Km and $\Delta L=1376.34$ Km. This  is much lower than $(2.17036)$ which is obtained for 
the experimental value of  $\theta _{13} =8.5^{\circ}$. So the presence of $\theta _{13}$ enhances the maximum violation of LGI by the amount $0.09274$ i.e.$4.6\%$. 
Increasing $\theta _{13}$ increases the maximum value of $C$ (fig.3).

(4)For the mass hierarchy parameter $\alpha = 0 $, \textit{i.e.} $m_{1}= m_{2}$ and the maximum value of $C$  is $2.09606$ for $\Delta L=1252.74$. Note that although now $m_{1}= m_{2}$ the maximum bound of $C$ is greater than $2$, i.e., we are still in the quantum domain. For two state neutrino oscillations, \cite{Gangopadhyay} $m_{1}= m_{2}$ implied that the maximum value of $C$ is $2$, i.e. \textit{e classical domain.!}.   $\alpha = 0.0305$ increases the the maximum value of the quantity $C$ by $3.7\%$ as shown in fig.4. 

(5)If $\delta _{CP} = 0$ in the expression for $C$ , the maximum value of $C$ reduces to $2.16553$ for $L_{1}=140.15$ Km and $\Delta L=1253.8$ Km. So presence of CP violating phase  parameter actually enhances the maximum violation of LGI by an amount $0.00483$ whic is $0.24\%$, a significant enhancement (fig.5). Compare this to the case of neutral kaon oscillations where including CP violation increased the maximum violation of LGI  by $0.008\%$ \cite{Gangopadhyay}. 

\begin{acknowledgements}
Animesh Sinha Roy thanks UGC-CSIR for providing a Research Fellowship Sr.No. 2061151173 under which this work was done.
\end{acknowledgements}

\section{Appendix}
\label{sec:6}
The details expression of the correlation function $C_{12}$ is given by
\begin{widetext}
\begin{eqnarray}
C_{12}&&=P_{\nu _{e},\nu_{e}}(L_{1},L_{2})-P_{\nu _{e},\nu_{\mu}}(L_{1},L_{2})-P_{\nu _{e},\nu_{\tau}}(L_{1},L_{2})
-P_{\nu _{\mu},\nu_{e}}(L_{1},L_{2})+P_{\nu _{\mu},\nu_{\mu}}(L_{1},L_{2})+P_{\nu _{\mu},\nu_{\tau}}(L_{1},L_{2})\nonumber\\
&&-P_{\nu _{\tau},\nu_{e}}(L_{1},L_{2})+P_{\nu _{\tau},\nu_{\mu}}(L_{1},L_{2})+P_{\nu _{\tau},\nu_{\tau}}(L_{1},L_{2})\nonumber\\
&&=\Bigg[1-\alpha ^{2}\sin ^{2}2\theta_{12}\frac{\sin ^{2}\big(\frac{VL_{1}}{2}\big)}{\Big(\frac{2EV}{\Delta m^{2}_{31}}\Big)^{2}}-4s^{2}_{13}\frac{\sin ^{2}\Big\{\Big(\frac{2EV}{\Delta m^{2}_{31}}-1\Big)\frac{\Delta m^{2}_{31}L_{1}}{4E}\Big\}}{\Big(\frac{2EV}{\Delta m^{2}_{31}}-1\Big)^{2}}\Bigg]\Bigg[1-2\alpha ^{2}\sin ^{2}2\theta_{12}\frac{\sin ^{2}\big(\frac{V(L_{2}-L_{1})}{2}\big)}{\Big(\frac{2EV}{\Delta m^{2}_{31}}\Big)^{2}}
 -8s^{2}_{13}\nonumber\\
 &&\times\frac{\sin ^{2}\Big\{\Big(\frac{2EV}{\Delta m^{2}_{31}}-1\Big)\frac{\Delta m^{2}_{31}(L_{2}-L_{1})}{4E}\Big\}}{\Big(\frac{2EV}{\Delta m^{2}_{31}}-1\Big)^{2}}\Bigg]-\Bigg[\alpha ^{2}\sin ^{2}2\theta_{12}c^{2}_{23}\frac{\sin ^{2}\big(\frac{VL_{1}}{2}\big)}{\Big(\frac{2EV}{\Delta m^{2}_{31}}\Big)^{2}}+4s^{2}_{13}s^{2}_{23}\frac{\sin ^{2}\Big\{\Big(\frac{2EV}{\Delta m^{2}_{31}}-1\Big)\frac{\Delta m^{2}_{31}L_{1}}{4E}\Big\}}{\Big(\frac{2EV}{\Delta m^{2}_{31}}-1\Big)^{2}} +2\alpha s_{13}\nonumber\\
&&\times \sin 2\theta_{12}\sin 2\theta_{23} \cos (\frac{\Delta m^{2}_{31}L_{1}}{4E} -\delta _{CP})\frac{\sin \big(\frac{VL_{1}}{2}\big)}{\Big(\frac{2EV}{\Delta m^{2}_{31}}\Big)}\frac{\sin \Big\{\Big(\frac{2EV}{\Delta m^{2}_{31}}-1\Big)\frac{\Delta m^{2}_{31}L_{1}}{4E}\Big\}}{\Big(\frac{2EV}{\Delta m^{2}_{31}}-1\Big)}\Bigg]\Bigg[2\alpha ^{2}\sin ^{2}2\theta_{12}c^{2}_{23}\frac{\sin ^{2}\big(\frac{V(L_{2}-L_{1})}{2}\big)}{\Big(\frac{2EV}{\Delta m^{2}_{31}}\Big)^{2}}\nonumber\\
&&+8s^{2}_{13}s^{2}_{23}\frac{\sin ^{2}\Big\{\Big(\frac{2EV}{\Delta m^{2}_{31}}-1\Big)\frac{\Delta m^{2}_{31}(L_{2}-L_{1})}{4E}\Big\}}{\Big(\frac{2EV}{\Delta m^{2}_{31}}-1\Big)^{2}}+4\alpha s_{13}\sin 2\theta_{12}\sin 2\theta_{23} \frac{\sin \big(\frac{V(L_{2}-L_{1})}{2}\big)}{\Big(\frac{2EV}{\Delta m^{2}_{31}}\Big)} \frac{\sin \Big\{\Big(\frac{2EV}{\Delta m^{2}_{31}}-1\Big)\frac{\Delta m^{2}_{31}(L_{2}-L_{1})}{4E}\Big\}}{\Big(\frac{2EV}{\Delta m^{2}_{31}}-1\Big)}\nonumber\\
&&\times \Bigg\{\cos (\frac{\Delta m^{2}_{31}(L_{2}-L_{1})}{4E} -\delta _{CP})-\sin \delta _{CP}\sin \Big(\frac{\Delta m^{2}_{31}(L_{2}-L_{1})}{4E}\Big)\Bigg\}-1\Bigg]-\Bigg[\alpha ^{2}\sin ^{2}2\theta_{12}s^{2}_{23}\frac{\sin ^{2}\big(\frac{VL_{1}}{2}\big)}{\Big(\frac{2EV}{\Delta m^{2}_{31}}\Big)^{2}}+4s^{2}_{13}c^{2}_{23}\nonumber\\
&&\times \frac{\sin ^{2}\Big\{\Big(\frac{2EV}{\Delta m^{2}_{31}}-1\Big)\frac{\Delta m^{2}_{31}L_{1}}{4E}\Big\}}{\Big(\frac{2EV}{\Delta m^{2}_{31}}-1\Big)^{2}}-2\alpha s_{13}\sin 2\theta_{12}\sin 2\theta_{23} \cos (\frac{\Delta m^{2}_{31}L_{1}}{4E} -\delta _{CP})\frac{\sin \big(\frac{VL_{1}}{2}\big)}{\Big(\frac{2EV}{\Delta m^{2}_{31}}\Big)}\frac{\sin \Big\{\Big(\frac{2EV}{\Delta m^{2}_{31}}-1\Big)\frac{\Delta m^{2}_{31}L_{1}}{4E}\Big\}}{\Big(\frac{2EV}{\Delta m^{2}_{31}}-1\Big)}\Bigg]\nonumber\\
&&\Bigg[2\alpha ^{2}\sin ^{2}2\theta_{12}s^{2}_{23}\frac{\sin ^{2}\big(\frac{V(L_{2}-L_{1})}{2}\big)}{\Big(\frac{2EV}{\Delta m^{2}_{31}}\Big)^{2}}+8s^{2}_{13}c^{2}_{23}\frac{\sin ^{2}\Big\{\Big(\frac{2EV}{\Delta m^{2}_{31}}-1\Big)\frac{\Delta m^{2}_{31}(L_{2}-L_{1})}{4E}\Big\}}{\Big(\frac{2EV}{\Delta m^{2}_{31}}-1\Big)^{2}}-4\alpha s_{13}\sin 2\theta_{12}\sin 2\theta_{23}\frac{\sin \big(\frac{V(L_{2}-L_{1})}{2}\big)}{\Big(\frac{2EV}{\Delta m^{2}_{31}}\Big)}\nonumber\\
&&\times \frac{\sin \Big\{\Big(\frac{2EV}{\Delta m^{2}_{31}}-1\Big)\frac{\Delta m^{2}_{31}(L_{2}-L_{1})}{4E}\Big\}}{\Big(\frac{2EV}{\Delta m^{2}_{31}}-1\Big)} \Bigg\{\cos (\frac{\Delta m^{2}_{31}(L_{2}-L_{1})}{4E} -\delta _{CP})-\sin \delta _{CP}\sin \Big(\frac{\Delta m^{2}_{31}(L_{2}-L_{1})}{4E}\Big)\Bigg\}-1\Bigg]
\end{eqnarray}
\end{widetext}

\end{document}